\documentstyle[12pt,epsfig]{article}
\newcommand{\nd}{\noindent}
\newcommand{\beq}{\begin{equation}}
\newcommand{\eeq}{\end{equation}}
\newcommand{\barr}{\begin{eqnarray}}
\newcommand{\earr}{\end{eqnarray}}
\newcommand{\ba}{\begin{array}}
\newcommand{\ea}{\end{array}}
\newcommand{\bfp}{\mbox{\boldmath $p$}}
\newcommand{\bfk}{\mbox{\boldmath $k$}}

\newcommand{\bfS}{\mbox{\boldmath $S$}}

\newcommand{\la}{\lambda}

\newcommand{\pup}{p^\uparrow}
\newcommand{\aup}{a^\uparrow}
\newcommand{\qup}{q^\uparrow}
\newcommand{\pdown}{p^\downarrow}
\newcommand{\adown}{a^\downarrow}
\newcommand{\qdown}{q^\downarrow}
\newcommand{\NP}[1]{{\it Nucl.\ Phys.}\ {\bf #1}}
\newcommand{\ZP}[1]{{\it Z.\ Phys.}\ {\bf #1}}
\newcommand{\PL}[1]{{\it Phys.\ Lett.}\ {\bf #1}}
\newcommand{\PR}[1]{{\it Phys.\ Rev.}\ {\bf #1}}
\newcommand{\PRL}[1]{{\it Phys.\ Rev.\ Lett.}\ {\bf #1}}

\def\lsim{\mathrel{\rlap{\lower4pt\hbox{\hskip1pt$\sim$}}\raise1pt\hbox{$<$}}}
\def\gsim{\mathrel{\rlap{\lower4pt\hbox{\hskip1pt$\sim$}}\raise1pt\hbox{$>$}}}
\def\nostrocostruttino#1\over#2{\mathrel{\mathop{\kern 0pt \rlap
{\hbox{$#1$}}} \hbox{\kern-.135em $#2$}}}

\begin{document}
%%%%%%%%%%%%%%%%%%%%%%%%%%%%%%%%%%%%%%%%%%%%%%%%%%%%%%%%%%%%%%%%%%%%%%%%%%%%%%
\begin{flushright}
DFTT 34/99 \\
VUTH 99-15 \\
INFNCA-TH9907 \\
hep-ph/9906418 \\
\end{flushright}
\vskip 1.5cm
\begin{center}
{\bf Predictions for single spin asymmetries in 
\mbox{\boldmath{$\ell p^{\uparrow} \!\! \to \pi X$}} 
and \mbox{\boldmath{$\gamma^* p^{\uparrow} \!\! \to \pi X$}}}\\
\vskip 0.8cm
{\sf M.\ Anselmino$^1$, M.\ Boglione$^2$, J.\ Hansson$^3$ and F.\ Murgia$^4$}
\vskip 0.5cm
{$^1$ Dipartimento di Fisica Teorica, Universit\`a di Torino and \\
      INFN, Sezione di Torino, Via P. Giuria 1, I-10125 Torino, Italy\\
\vskip 0.5cm
$^2$  Dept. of Physics and Astronomy, Vrije Universiteit Amsterdam, \\
De Boelelaan 1081, 1081 HV Amsterdam, The Netherlands \\ 
\vskip 0.5cm
$^3$ Department of Physics, Lule{\aa} University of Technology \\
SE-971 87 Lule{\aa}, Sweden \\
\vskip 0.5cm 
$^4$  Istituto Nazionale di Fisica Nucleare, Sezione di Cagliari\\
      and Dipartimento di Fisica, Universit\`a di Cagliari\\
      C.P. 170, I-09042 Monserrato (CA), Italy} \\
\end{center}
\vskip 1.5cm
\noindent
{\bf Abstract:} \\ 
Predictions for the single transverse spin asymmetry $A_N$ in semi-inclusive 
DIS processes are given; non negligible values of $A_N$ may arise from spin 
effects in the fragmentation of a polarized quark into a final hadron with 
a transverse momentum $\bfk_\perp$ with respect to the jet axis, the
so-called Collins effect. The elementary single spin asymmetry of
the fragmenting quark has been fixed in a previous paper, by fitting
data on $\pup p \to \pi X$, and the predictions given here are uniquely 
based on the assumption that the Collins effect is the only cause of
the observed single spin asymmetries in $\pup p \to \pi X$.  
Eventual spin and $\bfk_\perp$ dependences in quark distribution functions, 
the so-called Sivers effect, are also discussed.   
%%%%%%%%%%%%%%%%%%%%%%%%%%%%%%%%%%%%%%%%%%%%%%%%%%%%%%%%%%%%%%%%%%%%%%%%%%%%%%%
\newpage
%%%%%%%%%%%%%%%%%%%%%%%%%%%%%%%%%%%%%%%%%%%%%%%%%%%%%%%%%%%%%%%%%%%%%%%%%%%%%%%
\pagestyle{plain}
\setcounter{page}{1}
\nd
{\bf 1. Introduction}
\vskip 6pt

A phenomenological description of single spin asymmetries in the 
inclusive production of hadrons in high energy processes, within the 
realm of pQCD, seems now indeed possible after the extensive amount 
of work which has been recently done \cite{siv}-\cite{bm2}. 
Novel, hopefully universal, distribution and fragmentation functions
\cite{siv, col, noi1, qs3, boe} may give origin to sizeable single 
spin hadronic observables; such functions can be measured in some 
processes and then used to give theoretical predictions in other cases.

This phenomenological program, based on data on $\pup p \to \pi X$
and $\bar p^\uparrow p \to \pi X$ \cite{e704}, was initiated 
in Refs.~\cite{noi1, noi2} and continued in Ref.~\cite{noi3}; 
a similar attempt, together with a comprehensive theoretical discussion, 
is considered in Ref.~\cite{qs3}.   

Several possible origins of single spin asymmetries can be at work 
in $\pup p$ or $\bar p^\uparrow p$ processes: spin and intrinsic $\bfk_\perp$
dependences in the distribution functions of unpolarized quarks inside
transversely polarized nucleons \cite{siv, noi1}; spin and intrinsic 
$\bfk_\perp$ dependences in the fragmentation functions of transversely 
polarized quarks into pions or other hadrons \cite{col, noi3}; very recently  
also possible spin and intrinsic $\bfk_\perp$ dependences in the 
distribution functions of transversely polarized quarks inside unpolarized 
nucleons have been suggested \cite{boe}.    

The first two possibilities above have been considered respectively in 
Refs.~\cite{noi2} and \cite{noi3} and explicit expressions of the relevant 
elementary functions have been obtained: 
\beq
\Delta^N f_{a/p}(x_a,\bfk_{\perp a}) 
= \hat f_{a/\pup}(x_a, \bfk_{\perp a}) -  
\hat f_{a/\pdown}(x_a, \bfk_{\perp a})\,, \label{delf}
\eeq
{\it i.e.} the difference between the density numbers 
of partons $a$, with all possible polarization, 
longitudinal momentum fraction $x_a$ and intrinsic transverse momentum 
$\bfk_{\perp a}$, inside a transversely polarized proton, and 
\beq
\Delta^N D_{h/a}(z, \bfk_{\perp h}) = 
\hat D_{h/\aup}(z, \bfk_{\perp h}) - 
\hat D_{h/\adown}(z, \bfk_{\perp h})\,, \label{deld}
\eeq 
{\it i.e.} the difference between the density numbers
of hadrons $h$, with longitudinal momentum fraction $z$ and transverse 
momentum $\bfk_{\perp h}$ inside a jet originated by the fragmentation of a 
transversely polarized parton $a$. 

These two functions are strictly related to the functions denoted by 
$f_{1T}^\perp$ and $H_1^{\perp}$ in Refs.~\cite{bm1}, as it is
further discussed in Ref.~\cite{bm2}.  

We consider here the DIS processes $\ell \pup \to \pi X$ and 
$\gamma^* \pup \to \pi X$, with unpolarized leptons;
however, the same results hold in case of longitudinally polarized leptons,
it is only crucial for the proton spin to be orthogonal to the
scattering plane. As it was
remarked in Refs.~\cite {noi3} and \cite{alm} we do not expect
any spin effect in the nucleon distribution functions in such
DIS processes, because these should be suppressed by necessary initial
state interactions which mean higher powers of $\alpha_{em}$.
We remain then with the only possibility of the quark fragmentation
single spin asymmetry, $\Delta^ND_{h/a}(z, \bfk_\perp) \not= 0$,
the so-called Collins effect \cite{col}.

Such an effect was studied in Ref.~\cite{noi3} where, assuming that it is 
the only cause of the observed single spin asymmetry in $\pup p \to \pi X$,
an explicit expression of $\Delta^ND_{h/a}(z, \bfk_\perp)$, based
on a simple parametrization, was obtained. By using this expression here
we are able to give predictions for the single spin asymmetry
\beq
A_N = \frac{ d\sigma^{\uparrow} - d\sigma^{\downarrow} }
           { d\sigma^{\uparrow} + d\sigma^{\downarrow} }
    = \frac{ d\sigma^{\uparrow} - d\sigma^{\downarrow} }
           { 2 \, d\sigma^{unp} } \label{an}
\eeq
in DIS processes.

In the next Section we will consider the process $\ell \pup \to \pi X$,
in which the final lepton need not be detected, in complete analogy     
with the $\pup p \to \pi X$ process; parity invariance allows 
$A_N$ to be different from zero only if the nucleon spin has
a component perpendicular to the scattering plane. This is not the case 
for longitudinally polarized protons. However, also for such spin
configurations, one could have a single spin asymmetry by looking at 
the semi-inclusive process $\ell \pup \to \ell \pi X$ or 
$\gamma^* \pup \to \pi X$;
also a longitudinally polarized proton can have a non-zero spin component
perpendicular to the scattering plane, which is now the $\gamma^*$-$\pi$
plane. Predictions for such a process will be given in Section 3. 

In Section 4 we consider the hypothetical possibility that Sivers 
effect, {\it i.e.} the function $\Delta^N f$ of Eq.~(\ref{delf}), might be 
at work also in DIS processes \cite{dra}, despite the argument given above
about initial state interactions, and compute $A_N$ in $\ell \pup \to \pi X$
due to such an effect using the expressions of $\Delta^N f$ derived 
in Ref.~\cite{noi2}; we do this in order to compare the size and $x_F$ 
dependence of $A_N$ eventually originating from Sivers effect with those
of $A_N$ originating from Collins effect. 
Comments and conclusions are presented in Section 5.
   
\vskip 18pt
\nd
{\bf 2. Single spin asymmetry} \mbox{\boldmath{$A_N$}} {\bf for} 
\mbox{\boldmath{$\ell \pup \!\! \to \pi X$}}
\vskip 6pt

The single transverse spin asymmetry (\ref{an}) for the high energy process
$\ell \pup \to \pi X$ in which a final pion is detected 
with a large $p_T$ is given, assuming a straightforward generalization of the 
QCD factorization theorem \cite{col} to the case in which parton transverse 
motion is taken into account and considering $\bfk_\perp$ dependences
only in the fragmentation process, by \cite{noi3, alm}
\barr
& & 2 \, A_N \, \frac{ E_\pi \, d^3\sigma^{unp}}{d^{3} \bfp_\pi} =  
  \frac{E_\pi \, d^3\sigma^{\ell + \pup \to \pi + X}} {d^{3} \bfp_\pi}  
- \frac{E_\pi \, d^3\sigma^{\ell + \pdown \to \pi + X}} {d^{3} \bfp_\pi} 
\label{anc} \\
&=& \sum_q \int \frac {dx} {\pi z} \, d^2 \bfk_\perp \>    
\Delta_{_T} q(x) \> \Delta_{NN} \hat\sigma^q (x, \bfk_\perp) \,
\left[ \hat D_{\pi/q^\uparrow}(z, \bfk_\perp)
- \hat D_{\pi/q^\uparrow}(z, - \bfk_\perp) \right] \>, 
\nonumber
\earr
where $\Delta_{_T}q$ is the polarized number density for transversely spinning 
quarks $q$, also denoted by $h_1$ \cite{jaf}, and $\Delta_{NN} \hat\sigma^q$ 
is the elementary cross-section double spin asymmetry
\beq
\Delta_{NN} \hat\sigma^q = {d\hat \sigma^{\ell q^\uparrow \to 
\ell q^\uparrow} \over d\hat t} - {d\hat \sigma^{\ell q^\uparrow \to 
\ell q^\downarrow} \over d\hat t} \,\cdot
\label{del}
\eeq
Here $\uparrow$ and $\downarrow$ refer to directions ``up'' and ``down''
with respect to the scattering plane: $\uparrow$ is parallel to 
$\bfp_\ell \times \bfp_\pi$. Notice that Eq.~(\ref{anc}) holds
both for unpolarized and longitudinally polarized leptons.

Following the same procedure as in Ref.~\cite{noi3} we write Eq.~(\ref{anc})
as 
\barr
&&  \frac{E_\pi \, d^3\sigma^{\ell + \pup \to \pi + X}} {d^{3} \bfp_\pi}  
- \frac{E_\pi \, d^3\sigma^{\ell + \pdown \to \pi + X}} {d^{3} \bfp_\pi} = 
\label{d-d} \\
&=& \sum_q \int \frac {dx} {\pi z} \, d^2 \bfk_\perp \>    
P^{q/\pup} f_{q/p}(x) \, [\Delta_{NN} \hat\sigma^q (x, \bfk_\perp) 
- \Delta_{NN} \hat\sigma^q (x, - \bfk_\perp)] \>
\Delta^N D_{\pi/q}(z, \bfk_{\perp}) \nonumber
\earr
where we have used Eq.~(\ref{deld}) and where the integration on 
$\bfk_{\perp}$ now runs only over one half-plane of its components. 
We have also used $\Delta_{_T} q(x) = h_1(x) = P^{q/\pup} f_{q/p}(x)$ 
where $P^{q/\pup}$ is the polarization of the quark $q$ inside 
the transversely polarized proton $\pup$; it is twice the average value of 
the $\uparrow$ component of the quark spin. 

We now proceed as in Ref.~\cite{noi3} with the assumption that the 
dominant contribution is given by valence quarks, which is certainly
correct for large $x_F$ pions which originate from large $x$ quarks;
for this reason we also assume values of $P^{q/\pup}$ which do not depend 
on $x$. We take into account only the leading $\bfk_\perp$ contributions 
which originate from $\bfk_{\perp} = \bfk^{0}_{\perp}$ where
$\bfk^{0}_{\perp}$ lies in the overall scattering plane and its magnitude 
equals the average value of 
$\langle \bfk^{2}_{\perp}\rangle^{1/2}= k^{0}_{\perp}(z)$. That is, we have:
\barr
&&  \frac{E_\pi \, d^3\sigma^{\ell + \pup \to \pi + X}} {d^{3} \bfp_\pi}  
- \frac{E_\pi \, d^3\sigma^{\ell + \pdown \to \pi + X}} {d^{3} \bfp_\pi} 
\label{d-dapp} \\
&\simeq& \sum_q \int \frac {dx} {\pi z} \>    
P^{q/\pup} f_{q/p}(x) \, [\Delta_{NN} \hat\sigma^q (x, +\bfk_\perp^0) 
- \Delta_{NN} \hat\sigma^q (x, - \bfk_\perp^0)] \>
\Delta^N D_{\pi/q}(z, k_{\perp}^0) \>. \nonumber
\earr
The value of $z$ is fixed as function of $x$ and $\bfk_\perp$
by energy-momentum conservation in the elementary scattering. 

The average value $k_\perp^0$ of the transverse momentum 
of charged pions inside jets does in general depend on $z$; 
the $z$ dependence  was measured at LEP \cite{abr} and  
we use a fit to their data points \cite{noi3}: 
\beq
\frac{k_{\perp}^0(z)}{M} = 0.61 \; z^{0.27} \; (1-z)^{0.20}\;,
\label{kfit}
\eeq
with $M$ = 1 GeV/$c^2$.

The explicit expression of $\Delta^ND_{\pi/q}$ obtained in Ref.~\cite{noi3}
by fitting the data on $\pup p \to \pi X$ \cite{e704} is, for valence quarks:
\barr
\Delta^ND_{\pi^+/u}(z, k_\perp^0) &=& 
\Delta^ND_{\pi^-/d}(z, k_\perp^0) = \Delta^ND_{val}(z, k_\perp^0)
\nonumber \\ 
= 2 \Delta^ND_{\pi^0/u}(z, k_\perp^0) &=& 
2 \Delta^ND_{\pi^0/d}(z, k_\perp^0) = 
- \frac{k_{\perp}^0(z)}{M} \,
0.22 \, z^{2.33}\,(1-z)^{0.24} \,,
\label{delfit}
\earr
for $z \leq 0.97742$ and 
\beq
\Delta^ND_{val}(z, k_\perp^0) = - 2 D_{val}(z) = - 2 \times 1.102 \, z^{-1} 
(1-z)^{1.2} \quad \quad {\rm for} \quad z > 0.97742 \>. \label{bfit2}
\eeq
The values of $P^{u/\pup}$ and $P^{d/\pup}$, always from the fit of 
Ref.~\cite{noi3}, are 
\beq
P^{u/\pup}= \frac{2}{3} \quad \quad \quad P^{d/\pup}= - 0.88 \>;
\label{pud}
\eeq
notice that only their ratio $P_{u/d} = P^{u/\pup}/P^{d/\pup} = - 0.76$ has 
a physical meaning in that the overall normalization of the quark 
polarizations can always be absorbed by the overall normalization 
of $\Delta^ND_{\pi/q}$.  

As we commented in Ref.~\cite{noi3} both the expression of $\Delta^ND_{val}$ 
and the value of $P_{u/d}$ resulting from fitting the $\pup p \to \pi X$
data -- assuming spin and $\bfk_\perp$ effects only in the fragmentation 
process -- are somewhat surprising and a bit extreme: $\Delta^ND_{val}$
has to saturate at large $z$ the necessary inequality 
$|\Delta^ND_{val}(z, k_\perp^0)| \le 2D_{val}(z)$ and $P_{u/d}$ shows 
an opposite (as expected) but larger in magnitude (unexpected)
polarization of the $d$ valence quark with respect to the $u$ 
valence quark inside a polarized proton. It might be that relevant
or even dominant contributions to $A_N$ in $\pup p \to \pi X$ processes
originate from quark distribution functions, Eq.~(\ref{delf}), the so-called
Sivers effect \cite{siv, noi1, noi2}. 

Nevertheless, we assume here that Collins effect alone is responsible for
the observed single spin asymmetries and use Eqs.~(\ref{kfit})-(\ref{pud}) 
into Eq.~(\ref{d-dapp}) to obtain predictions
for $A_N$, Eq.~(\ref{an}). The unpolarized cross-section 
\beq
\frac{E_\pi \, d^3\sigma^{unp}}{d^{3} \bfp_\pi} =  
\sum_q \int \frac {dx} {\pi z} \> f_{q/p}(x) \> 
\frac{d\hat\sigma^{\ell q \to \ell q}}{d\hat t}(x) \>
D_{\pi/q}(z) \label{dsu}
\eeq
is computed using the same distribution and fragmentation functions as in 
Ref.~\cite{noi3}.

Our results are shown in Figs. 1-4, where we plot $A_N$ as a function
of $x_F$ at different centre of mass energies typical of experiments
running or planned at Jefferson Lab (CEBAF), DESY (HERMES), SLAC (E155) and 
CERN (SMC, COMPASS) respectively; $p_T$ is fixed and ranges, according 
to the different cases, from 1 to 2 GeV/$c$. The asymmetry turns out 
to be large and sizeable at large values of $x_F$ in all cases: we can 
unambiguously conclude that, if Collins effect is the main cause of
the single transverse spin asymmetries observed in 
$\pup p \to \pi X$ \cite{e704}, a similarly large asymmetry should be 
observed in lepto-production processes, $\ell \pup \to \pi X$.

Data on $A_N$ in such processes are not available yet, although they
might be soon obtainable from SMC and SLAC where DIS processes with 
transversely polarized protons are studied.
Protons are longitudinally polarized at HERMES and Jlab experiments
and parity invariance does not allow a single proton spin asymmetry in 
$\ell p^{\to} \to \pi X$; however, as we said in the introduction 
and will discuss in the next session, even in such a case one could
have a single transverse spin asymmetry looking at the $\gamma^* \pup
\to \pi X$ reaction.    
 
\vskip 18pt
\nd
{\bf 3. Single spin asymmetry} \mbox{\boldmath{$A_N$}} {\bf for} 
\mbox{\boldmath{$\gamma^* \pup \!\! \to \pi X$}}
\vskip 6pt

Let us consider now the double-inclusive process $\ell \pup \to \ell \pi X$
in which also the final lepton is detected, which allows to fix the
energy and momentum of the virtual photon exchanged between the
lepton and the nucleon; one can then consider the centre of mass process
$\gamma^* \pup \to \pi X$. A longitudinally polarized nucleon, {\it i.e.} 
a nucleon with its spin along the lepton direction in the laboratory
frame, may have a transverse spin component with respect to the 
$\gamma^*$-$\pi$ plane.

We fix our kinematics in Fig. 5; we have defined the  
$\gamma^* \pup \to \pi X$ scattering plane as the $xz$-plane;
the spin $\uparrow(\downarrow)$ is then along the $+y\,(-y)$
direction. The angle between the $\ell$-$\ell'$ and the $\gamma^*$-$\pi$
planes is $\beta$. We shall give values of the single spin asymmetry
$A_N$, Eq.~(\ref{an}), for this configuration. 

The analogue of Eq.~(\ref{anc}) is now
\barr
& & 2 \, A_N \, \frac{d\sigma^{\gamma^* p \to \pi X}}
{dx \, dQ^2 \, dz \, d^2p_T} 
= \frac{d\sigma^{\gamma^* \pup   \to \pi X}}{dx \, dQ^2 \, dz \, d^2p_T}
- \frac{d\sigma^{\gamma^* \pdown \to \pi X}}{dx \, dQ^2 \, dz \, d^2p_T}  
\label{ancg} \\
&=& \sum_q P^{q/\pup} f_{q/p}(x) 
\left[ \frac{d\hat\sigma^{\gamma^* \qup \to \qup}} {dQ^2}
- \frac{d\hat\sigma^{\gamma^* \qup \to \qdown}} {dQ^2} \right] \>
\left[ \hat D_{\pi/q^\uparrow}(z, \bfp_T)
- \hat D_{\pi/q^\uparrow}(z, - \bfp_T) \right] \>, 
\nonumber
\earr
where $x$ and $Q^2$ are the usual DIS variables and where the unpolarized
cross-section is:
\beq
\frac{d\sigma^{\gamma^* p \to \pi X}}{dx \, dQ^2 \, dz \, d^2p_T}
= \sum_q f_{q/p}(x) \> \frac{d\hat\sigma^{\gamma^* q \to q}} {dQ^2}
\> \hat D_{\pi/q}(z, \bfp_T) \>. \label{unpg}
\eeq

Notice that in the $\gamma^*$-$p$ c.m. frame, neglecting the transverse 
motion of the quarks inside the polarized proton -- that is, taking
into account only Collins effect -- the transverse momentum 
$\bfp_T$ of the produced pion is the same as the intrinsic $\bfk_\perp$
of the pion relative to fragmenting quark direction. The angle between the 
quark transverse spin direction and $\bfk_\perp$ -- sometimes referred to as 
Collins angle -- is fixed and equals $\pi/2$ in the configuration of Fig. 5.

The cross-section for the $\gamma^* q \to q$ process, with 
unpolarized quarks, is given by:
\barr
\frac{d\hat\sigma^{\gamma^* q \to q}}{dQ^2} &=& 
C \, \frac{1}{2} \sum_{\la^{\,}_q, \la_{q'}} 
\sum_{\la^{\,}_\gamma, \la^\prime_{\gamma}}
M_{\la_{q'}; \la^{\,}_\gamma, \la^{\,}_q} \>
M^*_{\la_{q'}; \la^\prime_{\gamma}, \la^{\,}_q} \>
\rho_{\la^{\,}_\gamma, \la^\prime_{\gamma}}(\gamma^*) \label{gqqu} \\
&\equiv& C \, \frac{1}{2} \sum_{\la^{\,}_q, \la_{q'}} 
\sum_{\la^{\,}_\gamma, \la^\prime_{\gamma}}
\langle \la_{q'} | \la^{\,}_\gamma, \la^{\,}_q \rangle \>
\langle \la_{q'} | \la^\prime_{\gamma}, \la^{\,}_q \rangle^* \>
\rho_{\la^{\,}_\gamma, \la^\prime_{\gamma}}(\gamma^*) \>, \nonumber 
\earr
where the $M$'s are the helicity amplitudes for the $\gamma^* q \to q$
process, $C$ contains kinematical and flux factors
which cancel out in the ratio giving $A_N$ and  
$\rho(\gamma^*)$ is the helicity density matrix of the virtual
photon emitted by the lepton; its explicit expression depends on the
DIS variables and can be found, for example, in Ref.~\cite{abmp}.
Analogously one has, for transversely polarized quarks:
\beq
\frac{d\hat\sigma^{\gamma^* \qup \to q^{\uparrow (\downarrow)}}}{dQ^2} = 
C \, \sum_{\la^{\,}_\gamma, \la^\prime_\gamma}
\langle \uparrow (\downarrow) | \la^{\,}_\gamma, \uparrow \rangle \>
\langle \uparrow (\downarrow) | \la^\prime_\gamma, \uparrow \rangle^* \>
\rho_{\la^{\,}_\gamma, \la^\prime_{\gamma}}(\gamma^*) \>. \label{gqqp}
\eeq

In terms of helicity states the transverse spins ``up'' and ``down''
are given, for the initial ($q$) and final ($q'$) quarks, by:
\beq
|q; \uparrow \rangle = \frac{1}{\sqrt 2} \> \left( |+\rangle -i |-\rangle
\right) \quad\quad
|q'; \uparrow, \downarrow \rangle = \frac{1}{\sqrt 2} \> 
\left( \pm |+\rangle +i |-\rangle \right) \>.
\label{thel}
\eeq

One finds:
\beq
\frac{d\hat\sigma^{\gamma^* q \to q}}{dQ^2} = C \, e^2 \, e_q^2 \, Q^2 \>
\frac{1 + (1-y)^2}{(2-y)^2}
\label{gqque}
\eeq
and
\beq
\frac{d\hat\sigma^{\gamma^* \qup \to q^\uparrow}}{dQ^2} - 
\frac{d\hat\sigma^{\gamma^* \qup \to q^\downarrow}}{dQ^2} =
2 \, C \, e^2 \, e_q^2 \, Q^2 \frac{1-y}{(2-y)^2} \> \cos2\beta
\label{gqqpe}
\eeq 
where $y = Q^2/xs$; notice that a dependence on the angle $\beta$ 
(see Fig. 5) remains in the elementary polarized process.

Insertion of Eqs.~(\ref{gqque}) and (\ref{gqqpe}) into 
Eqs.~(\ref{ancg}) and (\ref{unpg}) obtains:
\beq 
A_N^{\gamma^* \pup \to \pi X}(x, Q^2, z, p_T) = 
\frac{(1-y) \, \cos2\beta}{[1 + (1-y)^2]} \>
\frac{\sum_q e_q^2 \, P^{q/\pup} \, f_{q/p}(x) \> \Delta^ND_{\pi/q}(z, p_T)}
{\sum_q e_q^2 \, f_{q/p}(x) \> D_{\pi/q}(z)} \> \cdot \label{ancgr}
\eeq
We give numerical estimates of $A_N/|\cos2\beta|$ using $\Delta^ND_{\pi/q}$ 
and $P^{q/\pup}$ from Eqs.~(\ref{delfit})-(\ref{pud}).
The average $p_T$ value is given, at any $z$, by Eq.~(\ref{kfit});
the $p_T$ dependence in the fragmentation function in the denominator of
Eq.~(\ref{ancgr}) is neglected.  

The actual asymmetry $A_N$ measurable in processes initiated by 
longitudinally polarized protons -- {\it i.e.} protons with spin parallel
to the lepton direction -- is smaller than that given in Eq. (\ref{ancgr}),
which assumes a $P = 100\%$ nucleon polarization perpendicular to the 
$\gamma^*$-$\pi$ scattering plane. The real kinematical situation is shown 
in Fig. 6, from which one sees that the observed asymmetry is given by
Eq. (\ref{ancgr}) multiplied by $P\sin\theta_\gamma \sin\Phi$ (the 
component of the actual target polarization $P$ perpendicular to the 
production plane) and in which one sets $\beta = \Phi$. Some preliminary
data from HERMES and SMC show the $\Phi$ dependence of $A_N$
\cite{herm,smc}.  

In Figs. 7-8 we plot the value of $A_N(x, Q^2, z)$ for CEBAF and  
HERMES planned or running energies, as a function of $z$, for several values 
of $x$ and at fixed values of $Q^2$. These plots show that the asymmetry is  
large, for large $x$ values, but only at very large $z$ values, which 
might be difficult to reach experimentally. 

The $z$ dependence of $A_N$ is dominated by the $z$ dependence of  
$\Delta^ND_{\pi/q}/D_{\pi/q}$; actually, in all cases in which only one 
flavour $q$ contributes dominantly, Eq.~(\ref{ancgr}) simplifies to
\beq 
A_N^{\gamma^* \pup \to \pi X}(y, z, p_T) \simeq 
\frac{(1-y) \, \cos2\beta}{[1 + (1-y)^2]} \>
\frac{P^{q/\pup} \, \Delta^ND_{\pi/q}(z, p_T)}
{D_{\pi/q}(z)} \> \cdot \label{ancgrs}
\eeq
It is then clear why $A_N$ is small at small $z$ values and is large 
and constant for $z \ge 0.977$: this is due to the behaviour of
$\Delta^ND_{val}$ resulting from fitting the $\pup p \to \pi X$ data,
see Fig. 3 of Ref.~\cite{noi3}. In particular the saturated behaviour 
of $\Delta^ND_{val}$ at large $z$, Eq.~(\ref{bfit2}), forces the large
value of $A_N$ when $z \to 1$. All these are typical features of the 
Collins effect, assuming that it is the only origin of single
transverse spin asymmetries in $\pup p \to \pi X$ processes, and it
would be very interesting to  check them experimentally.
  
\vskip 18pt
\nd
{\bf 4. Single spin asymmetry} \mbox{\boldmath{$A_N$}} {\bf for} 
\mbox{\boldmath{$\ell \pup \!\! \to \pi X$}} {\bf with Sivers effect}
\vskip 6pt

Sivers effect is forbidden by time reversal invariance 
for free particles \cite{col}; however, one can avoid this conclusion by 
invoking some soft initial state interactions between the incoming 
particles \cite{noi1, boe}. This is plausible for $pp$ interactions or, 
in general, hadron-hadron interactions where many soft gluons might 
be exchanged, or some external gluonic fields should be present;
the assumption of the validity of the factorization theorem also in this 
case allows then to compute $A_N$ due to Sivers effect \cite{noi1, noi2}.
  
We cannot use such an argument in $\ell N$ processes because 
initial state interactions would now lead to strong suppressions of 
${\cal O}(\alpha_{em})$. It has been suggested \cite{dra} that 
some spin-isospin interactions might avoid the time reversal 
problem in chiral models, with quarks moving in a background of chiral 
fields, with a spin-isospin interaction which mixes states of different 
flavours under time reversal.

We do not claim here the definite validity of the above suggestion, which 
would require further developments and for which we have no explicit 
model; we simply assume that Sivers effect can be at work also
in $\ell p$ interactions and consistently compute $A_N$ in  
$\ell \pup \! \to \pi X$ processes due solely to this origin. 
Our aim is that of comparing the values and features of $A_N$ due 
to Sivers effect to the values and features of $A_N$ due to Collins 
effects, as computed in Section 1, in order to eventually distinguish 
between the two cases, independently of their theoretical plausibilities.    

If we assume that only Sivers effect is at work, Eq.~(\ref{anc}) changes
into: 
\barr
& & 2 \, A_N \, \frac{ E_\pi \, d^3\sigma^{unp}}{d^{3} \bfp_\pi} =  
  \frac{E_\pi \, d^3\sigma^{\ell + \pup \to \pi + X}} {d^{3} \bfp_\pi}  
- \frac{E_\pi \, d^3\sigma^{\ell + \pdown \to \pi + X}} {d^{3} \bfp_\pi} 
\label{ans} \\
&=& \sum_q \int \frac {dx} {\pi z} \, d^2 \bfk_\perp \>    
\Delta^Nf_{q/p}(x, \bfk_\perp) \> 
\frac{d\hat\sigma^{\ell q \to \ell q}}{d\hat t}(x, \bfk_\perp\>) \> 
D_{\pi/q}(z) \nonumber
\earr
where $d\sigma^{unp}$ is the same as in Eq.~(\ref{dsu}). Again,
Eq.~(\ref{ans}) holds both for unpolarized and longitudinally polarized 
leptons: the proton spin has to be orthogonal to the scattering plane. 

We proceed in the same way as in Section 1, and Refs.~\cite{noi1, noi2}, to
obtain, in analogy to Eq.~(\ref{d-dapp}): 
\barr
&&  \frac{E_\pi \, d^3\sigma^{\ell + \pup \to \pi + X}} {d^{3} \bfp_\pi}  
- \frac{E_\pi \, d^3\sigma^{\ell + \pdown \to \pi + X}} {d^{3} \bfp_\pi} 
\label{d-dapps} \\
&\simeq& \sum_q \int \frac {dx} {\pi z} \>    
\Delta^Nf_{q/p}(x, k^0_\perp) \,  
[ d\hat\sigma^{\ell q \to \ell q}(x, + \bfk_\perp^0) 
- d\hat\sigma^{\ell q \to \ell q}(x, - \bfk_\perp^0)] \>
D_{\pi/q}(z) \>. \nonumber
\earr
The function $\Delta^Nf_{q/p}(x, k^0_\perp)$ can be found in 
Refs.~\cite{noi2}, where it was derived for $u$ and $d$ valence quarks
by fitting the data on $\pup p \to \pi X$ (notice that, as
commented in Ref.~\cite{noi3}, the values of $N_a$ given in Eq. (9)
of Ref.~\cite{noi2} should be multiplied by a factor 4):
\barr
\Delta^Nf_{u/p}(x, k_\perp^0) &=& \frac{k^0_\perp(x)}{M} \>
14.72 \, x^{1.34} (1-x)^{3.58} \label{delfu} \\
\Delta^Nf_{d/p}(x, k_\perp^0) &=& - \frac{k^0_\perp(x)}{M} \>
4.96 \, x^{0.76} (1-x)^{4.14} \label{delfd} 
\earr
with
\beq
\frac{k^0_\perp(x)}{M} = 0.47 \, x^{0.68} (1-x)^{0.48} \>. \label{kts}
\eeq
  
{}From Eqs.~(\ref{ans})-(\ref{kts}) and (\ref{dsu}) we can compute $A_N$ 
as hypothetically given by Sivers effect alone; the results, at the same
energy and $p_T$ values of the similar results obtained from Collins
effects, Figs. 1-4, are shown in Figs. 9-12.

Also in this case $A_N$ turns out to be large and detectable. The $x_F$
dependences and the maximum values are clearly different from those
generated by Collins effect: $A_N$ is smoother and less pronounced 
for $\pi^+$ and $\pi^0$ and steeper and larger for $\pi^-$, the opposite 
of what is shown in Figs. 1-4 where $\pi^+$ and $\pi^0$ have a stronger 
dependence on $x_F$ and higher absolute values than $\pi^-$. In particular   
Sivers effect would produce the biggest (in magnitude) asymmetries for
$\pi^-$'s, whereas Collins effect does it for $\pi^+$'s.   

\vskip 18pt
\nd
{\bf 5. Comments and conclusions}
\vskip 6pt

Single transverse spin asymmetries have been observed in $\pup p \to \pi X$
and $\bar p^\uparrow p \to \pi X$ processes \cite{e704} in a kinematical 
region where the unpolarized cross-section is well described by pQCD and
the factorization theorem at leading twist: these measured large values of 
$A_N$ are very surprising because the same pQCD and factorization theorem 
at leading twist predict negligible values of $A_N$. They have to be 
understood and checked with further experiments.

Several possible explanations have been suggested within QCD 
\cite{siv}-\cite{bm2} or even classical models \cite{men}, and indeed 
phenomenological consistent descriptions of single spin asymmetries in
several processes seem now possible and should be pursued in order to
understand the basic underlying mechanisms which originate them.
We have studied two possible mechanisms \cite{noi1}-\cite{noi3}, the so-called 
Sivers and Collins effects, respectively described by the new $\bfk_\perp$ 
and spin dependent functions of Eqs.~(\ref{delf}) and (\ref{deld}). 
By fitting the available data we have deduced simple explicit expressions for 
$\Delta^Nf_{q/p}(x, k^0_\perp)$ and $\Delta^ND_{\pi/q}(z, k^0_\perp)$.

We have considered here the inclusive production of pions in lepton-nucleon
interactions in the process $\ell \pup \to \pi X$ in complete analogy 
to $\pup p \to \pi X$. Independently of the details of the proposed mechanisms 
to generate single spin asymmetries, the mere fact of measuring $A_N$ 
in such a process is significant and important. As we said, for several 
reasons we expect that single spin asymmetry in $\ell N$ interactions     
can only originate from single spin effects in the fragmentation of a 
transversely polarized quark into the final pion; if this is true
also for $\pup p \to \pi X$, as some authors think, then $A_N$ should be of
similar nature in $\ell \pup$ and $p \pup$ (or $\pup p$) interactions. 
The different elementary dynamics in the two cases -- $\ell q \to \ell q$
versus QCD dynamics like $qq \to qq$ or $qg \to qg$ -- should result
in (very interesting) differences in the shape of $A_N$, but not in the 
size of its magnitude.  

In Figs. 1-4 we have presented our predictions for $A_N$ in 
$\ell \pup \to \pi X$ processes 
exactly under the above assumption: that only Collins effect can be 
responsible for some single transverse spin dependence. If data agree 
with our predictions it would confirm this assumption and the consistency
of our phenomenological approach to the computations of inclusive single 
spin asymmetries; if not, it might be that other effects, like Sivers effect,
are responsible for $A_N$ in hadron-hadron collisions. Data might already 
be available from SMC and SLAC experiments, where nucleons can be 
transversely polarized, and will be available in the near future from
other experiments which have now only longitudinally polarized nucleons.      

Single transverse spin asymmetries may be measurable also in the 
case of longitudinally polarized nucleons provided one looks at the double 
inclusive process, $\ell \pup \to \ell \pi X$ from which one can reconstruct 
the $\gamma^* \pup \to \pi X$ reaction, which, in general, occurs in 
a plane different from the $\ell$-$\ell'$ plane where the longitudinal 
nucleon spin lies (Fig. 5). This further selection of events - to those
such that the proton polarization vector has a significant component
perpendicular to the $\gamma^*$-$\pi$ scattering plane - might greatly
reduce the experimental available information; also, our predictions,
Figs. 7-8, show sizeable values of $A_N$ only at large $z$ values, which 
might be difficult to reach.

Although Sivers effect is not expected to contribute to $A_N$ in 
lepton-nucleon interactions, unless some really new and important mechanism 
allows avoiding problems with time reversal invariance, we have computed 
$A_N$ assuming only Sivers effect to be at work, Figs. 9-12; a comparison 
with the corresponding results originated by Collins effect, Figs. 1-4, shows 
interesting differences: in particular $|A_N|$ is biggest for $\pi^+$ 
according to Collins mechanism, whereas it is biggest for $\pi^-$ according
to Sivers.  
  
To conclude, we believe that the measurement of $A_N$ -- possibly in the 
simple and more direct channel $\ell \pup \to \pi X$ -- can and should 
be performed by several experiments; the comparison with existing values of 
$A_N$ from $\pup p \!\! \to \pi X$ and $\bar p^\uparrow p \to \pi X$ 
processes would immediately allow to draw conclusions on possible origins 
of single spin effects. Once more, spin dependent measurements 
probe the hadronic structure and the corresponding theories and models
at deeper levels than usual unpolarized quantities. 

\vskip 18pt
\noindent
{\bf Acknowledgements}
\vskip 6pt
This work has been supported by the European Community under contract 
CHRX-CT94-0450.

%%%%%%%%%%%%%%%%%%%%%%%%%%%%%%%%%%%%%%%%%%%%%%%%%%%%%%%%%%%%%%%%%%%%

\newpage

%%%%%%%%%%%%%%%%%%%%%%%%%% Fig. 1 %%%%%%%%%%%%%%%%%%%%%%%%%%%%%%%%%%%%%%%%%
\begin{figure}[t]
\begin{center}
\mbox{~\epsfig{file=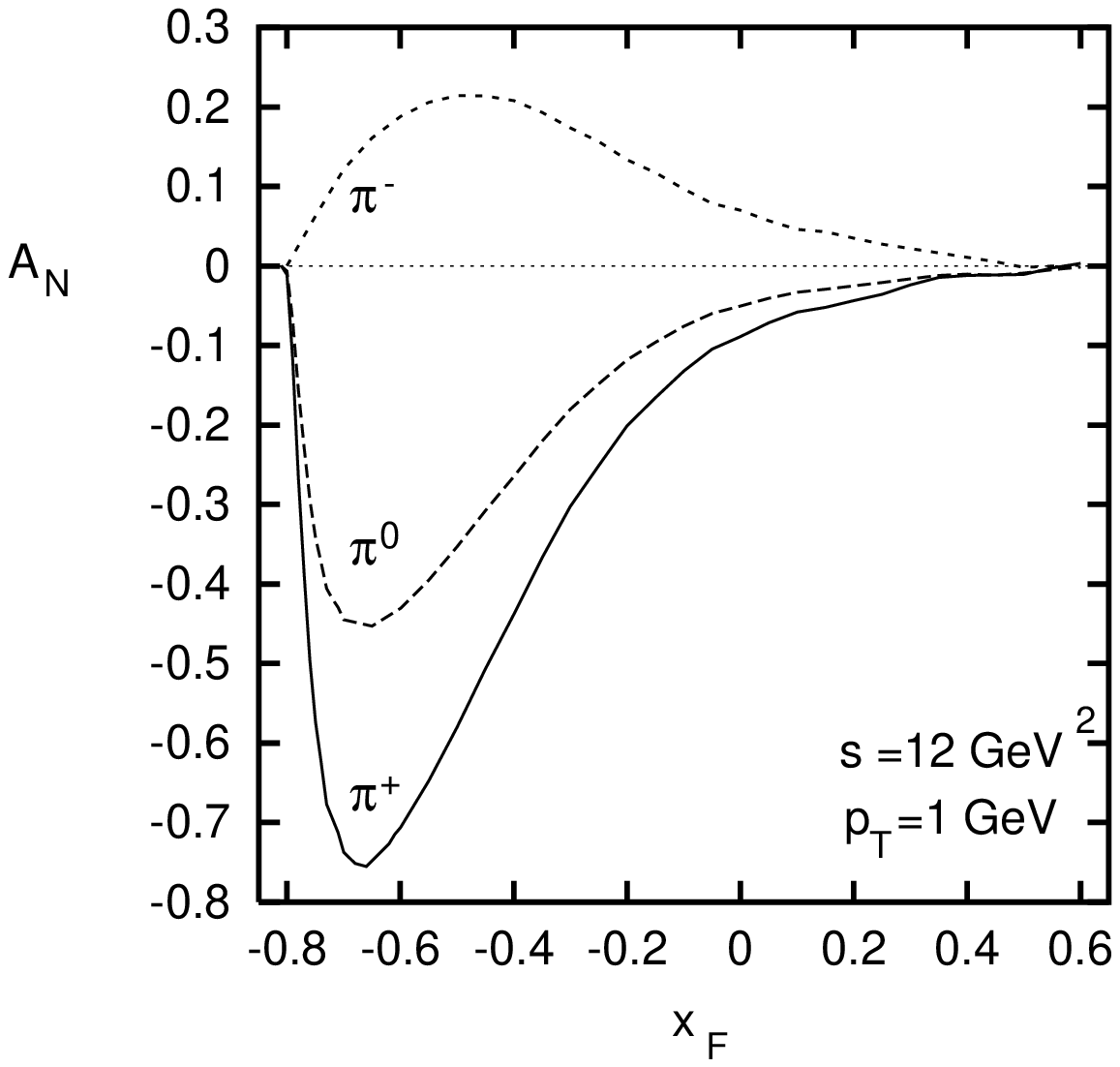,angle=-90,width=12cm}}
\vspace{1cm}
\caption{Predictions for the single spin asymmetry $A_N$ in the DIS process 
$l p^{\uparrow} \to \pi X$ as a function of $x_F$: the solid line refers 
to $\pi^+$, the dashed line to $\pi^0$ and the dotted line to $\pi^-$.
We assume that only Collins effect is active and
the function $\Delta^N D(z,k_\perp)$ needed to calculate $A_N$ is derived from 
a fit to E704 experimental data on $p^{\uparrow} p \to \pi X$
\protect\cite{noi3}.
The value of $s$ is chosen as a planned CEBAF value, $s=12$ GeV$^2$, 
and the transverse momentum of the pion is fixed to $p_T=1.0$ GeV.}
\end{center}
\end{figure}
%%%%%%%%%%%%%%%%%%%%%%%%%%%%%%%%%%%%%%%%%%%%%%%%%%%%%%%%%%%%%%%%%%%%%%%%%%%

\newpage

%%%%%%%%%%%%%%%%%%%%%%%%%%% Fig. 2 %%%%%%%%%%%%%%%%%%%%%%%%%%%%%%%%%%%%%%%%
\begin{figure}[t]
\begin{center}
\mbox{~\epsfig{file=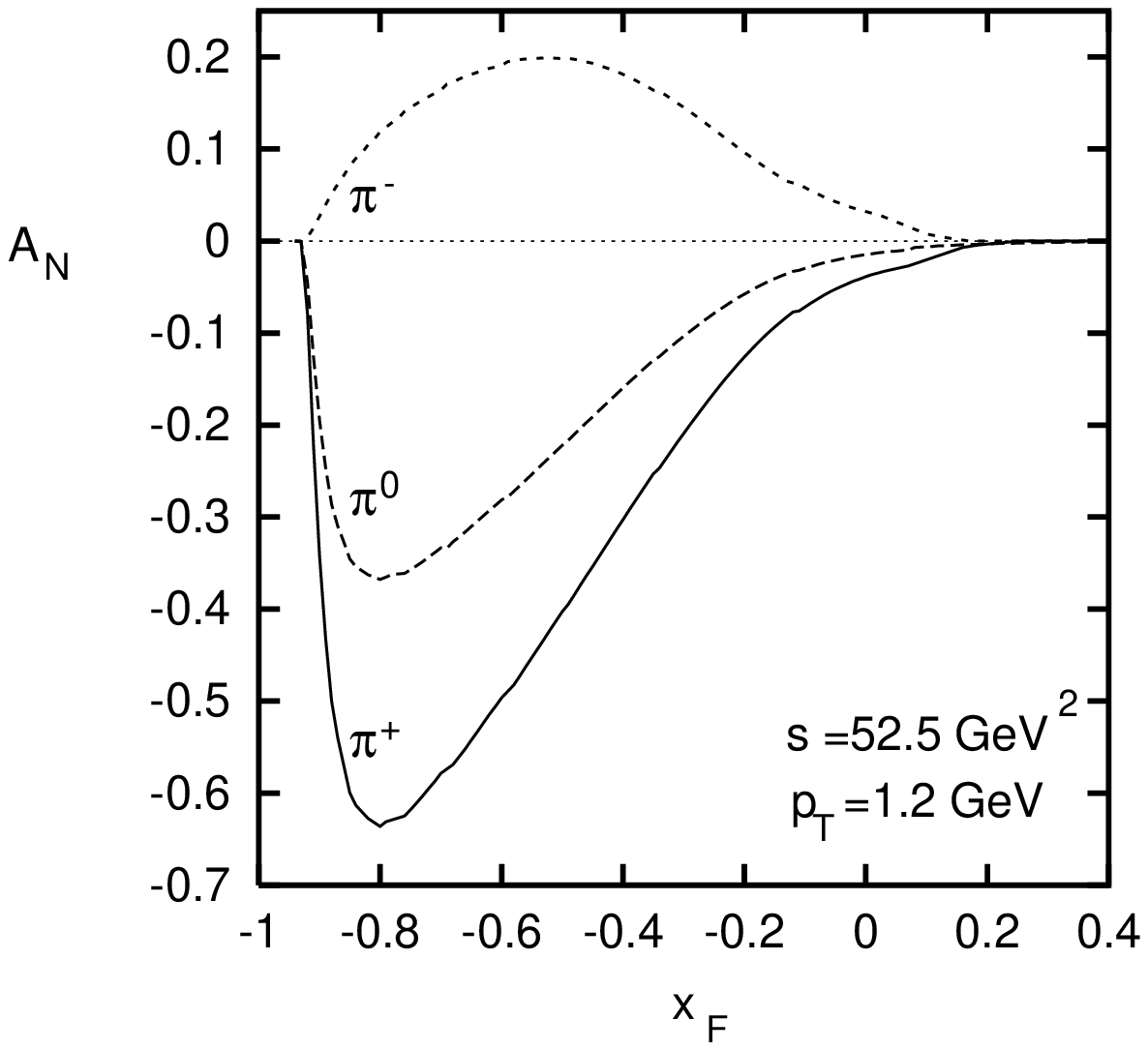,angle=-90,width=12cm}}
\vspace{1cm}
\caption{The same as in Fig.~1 for typical
%Single spin asymmetry for pion production in the DIS process 
%$\l  p^{\uparrow} \to \pi X$ as a function of $x_F$.
%The solid line refers to $\pi^+$, the dashed line to 
%\pi^0$ and the dotted line to $\pi^-$.
%the function $\Delta^N D(z,k_\perp)$ needed to calculate $A_N$ is taken from 
%the fit on E704 experimental data on $p^{\uparrow}  p \to \pi X$. 
%the value of $s$ is set to be the 
kinematical values of HERMES experiment, $s=52.5$ GeV$^2$, and $p_T=1.2$ GeV.}
\end{center}
\end{figure}
%%%%%%%%%%%%%%%%%%%%%%%%%%%%%%%%%%%%%%%%%%%%%%%%%%%%%%%%%%%%%%%%%%%%%%%%%%%

\newpage

%%%%%%%%%%%%%%%%%%%%%%%%%%%% Fig. 3 %%%%%%%%%%%%%%%%%%%%%%%%%%%%%%%%%%%%%%%
\begin{figure}[t]
\begin{center}
\mbox{~\epsfig{file=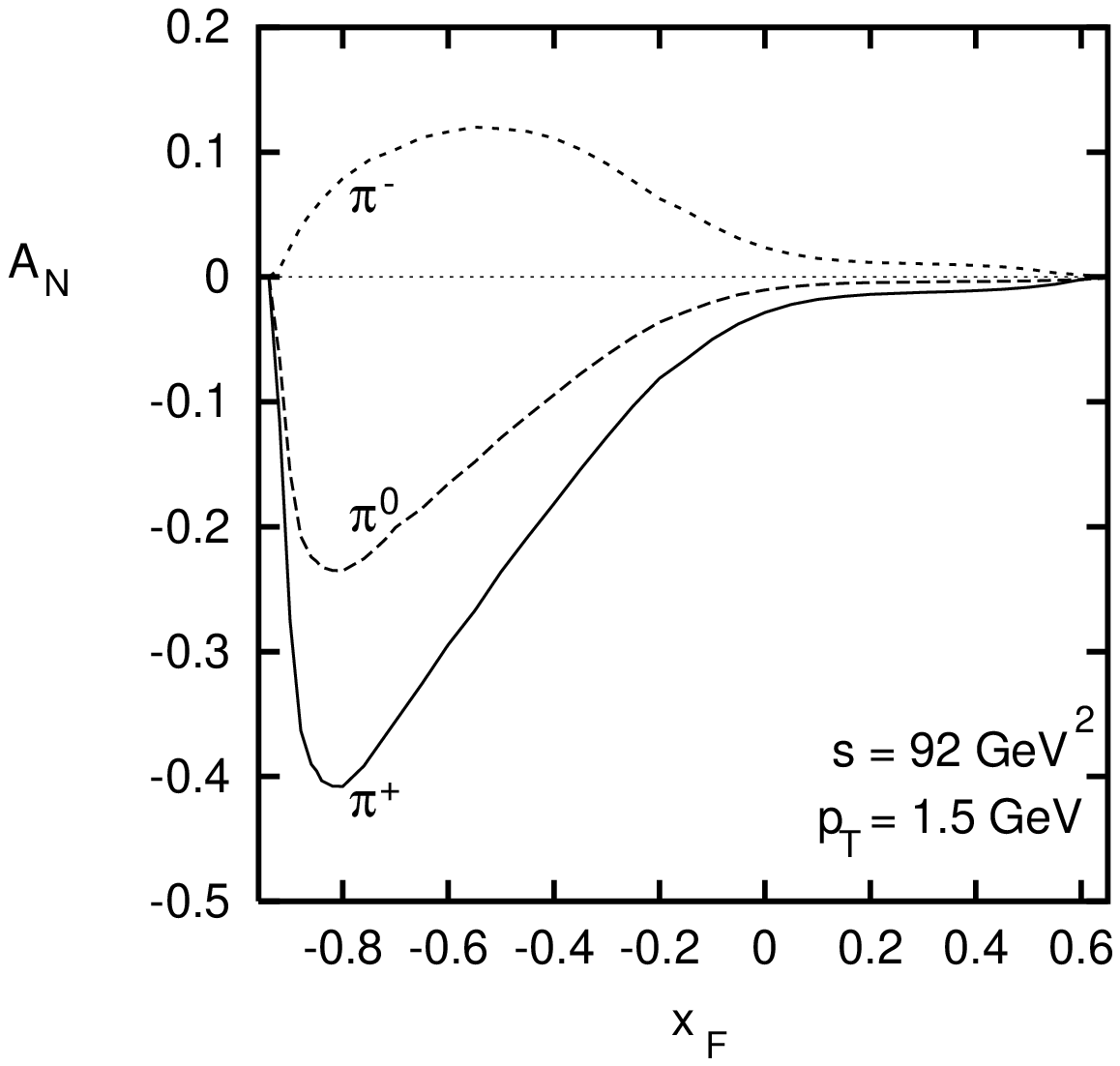,angle=-90,width=12cm}}
\vspace{1cm}
\caption{The same as in Fig.~1 
%Single spin asymmetry for pion production in the DIS process 
%$l  p^{\uparrow} \to \pi X$ as a function of $x_F$.
% The solid line refers to $\pi^+$, the dashed line to 
%$\pi^0$ and the dotted line to $\pi^-$. 
%Notice that here only Collins effects are present. 
%The value of $s$ is set to be 
at SLAC energy, $s=92$ GeV$^2$, with the 
transverse momentum of the pion fixed to $p_T=1.5$ GeV.}
\end{center}
\end{figure}
%%%%%%%%%%%%%%%%%%%%%%%%%%%%%%%%%%%%%%%%%%%%%%%%%%%%%%%%%%%%%%%%%%%%%%%%%%%

\newpage

%%%%%%%%%%%%%%%%%%%%%%%%%%%%% Fig. 4 %%%%%%%%%%%%%%%%%%%%%%%%%%%%%%%%%%%%%%
\begin{figure}[t]
\begin{center}
\mbox{~\epsfig{file=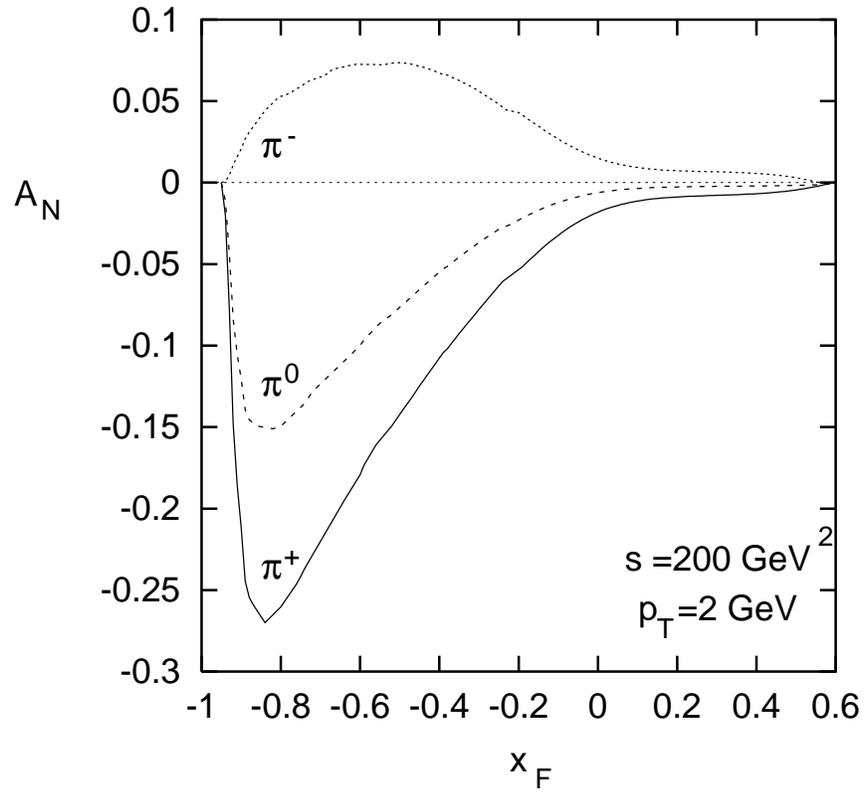,angle=-90,width=12cm}}
\vspace{1cm}
\caption{The same as in Fig.~1
%Single spin asymmetry for pion production in the DIS process 
%$l  p^{\uparrow} \to \pi X$ as a function of $x_F$.
% The solid line refers to $\pi^+$, the dashed line to 
%$\pi^0$ and the dotted line to $\pi^-$. 
%Notice that here only Collins effects are present. 
%The value of $s$ is set to be the 
for the SMC energy value $s=200$ GeV$^2$, and a  
transverse momentum of the pion $p_T=2$ GeV. Notice that 
$s=400$ GeV$^2$ would yield a very similar result.}
\end{center}
\end{figure}
%%%%%%%%%%%%%%%%%%%%%%%%%%%%%%%%%%%%%%%%%%%%%%%%%%%%%%%%%%%%%%%%%%%%%%%%%%%

\newpage

%%%%%%%%%%%%%%%%%%%%%%%%%%%%% Fig. 5 %%%%%%%%%%%%%%%%%%%%%%%%%%%%%%%%%%%%%%
\begin{figure}[t]
\begin{center}
\mbox{~\epsfig{file=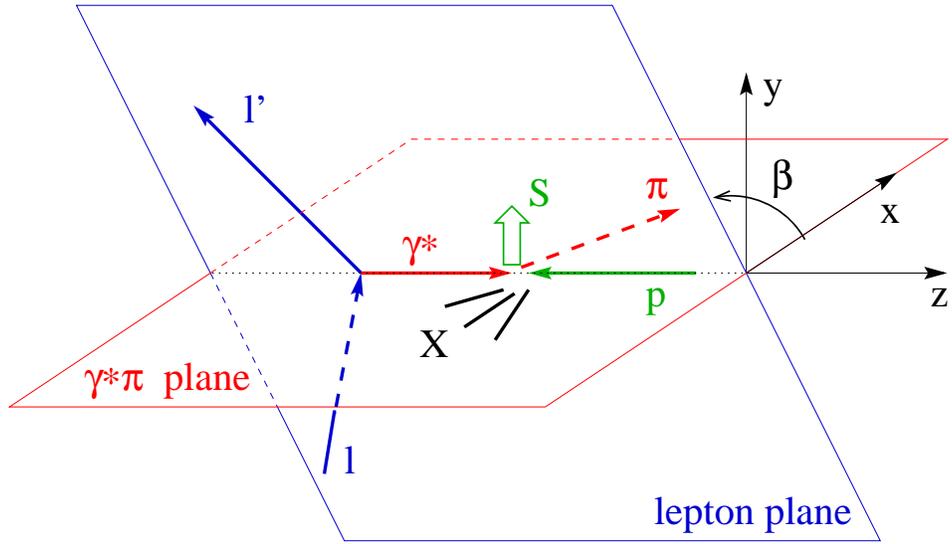,angle=0,width=14cm}}
\vspace{1cm}
\caption{Kinematical and spin configurations adopted for the computation
of $A_N$ in $\gamma^* \pup \to \pi X$ processes, Eq. (\ref{ancgr}) of text.}
\end{center}
\end{figure}
%%%%%%%%%%%%%%%%%%%%%%%%%%%%%%%%%%%%%%%%%%%%%%%%%%%%%%%%%%%%%%%%%%%%%%%%%%% 

\newpage

%%%%%%%%%%%%%%%%%%%%%%%%%%%%% Fig. 6 %%%%%%%%%%%%%%%%%%%%%%%%%%%%%%%%%%%%%%
\begin{figure}[t]
\begin{center}
\mbox{~\epsfig{file=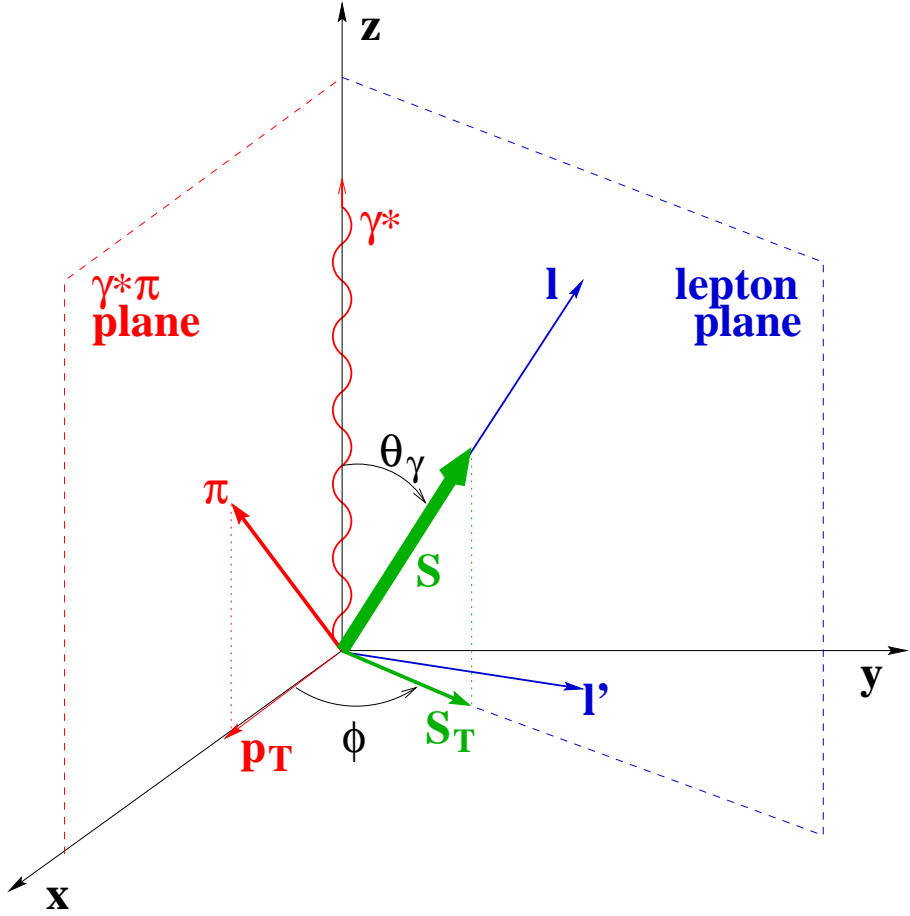,angle=0,width=12cm}}
\vspace{1cm}
\caption{The planes $\gamma^*$-$\pi$ and $\ell$-$\ell'$ in the reference
frame where the $\gamma^*$ moves along the $z$-axis and the nucleon  
at rest is longitudinally polarized, {\it i.e.} its spin is parallel to 
the $\ell$-direction of motion. The nucleon spin component perpendicular
to the $\gamma^*$-$\pi$ plane is $S \sin\theta_\gamma \sin\Phi =
S_T \sin\Phi$ and the Collins angle between $\bfS_T$ and $\bfp_T$ is $\Phi$.} 
\end{center}
\end{figure}
%%%%%%%%%%%%%%%%%%%%%%%%%%%%%%%%%%%%%%%%%%%%%%%%%%%%%%%%%%%%%%%%%%%%%%%%%%% 

\newpage

%%%%%%%%%%%%%%%%%%%%%%%%%%%%%%% Fig. 7 %%%%%%%%%%%%%%%%%%%%%%%%%%%%%%%%%%%%
\begin{figure}[t]
\begin{center}
\mbox{~\epsfig{file=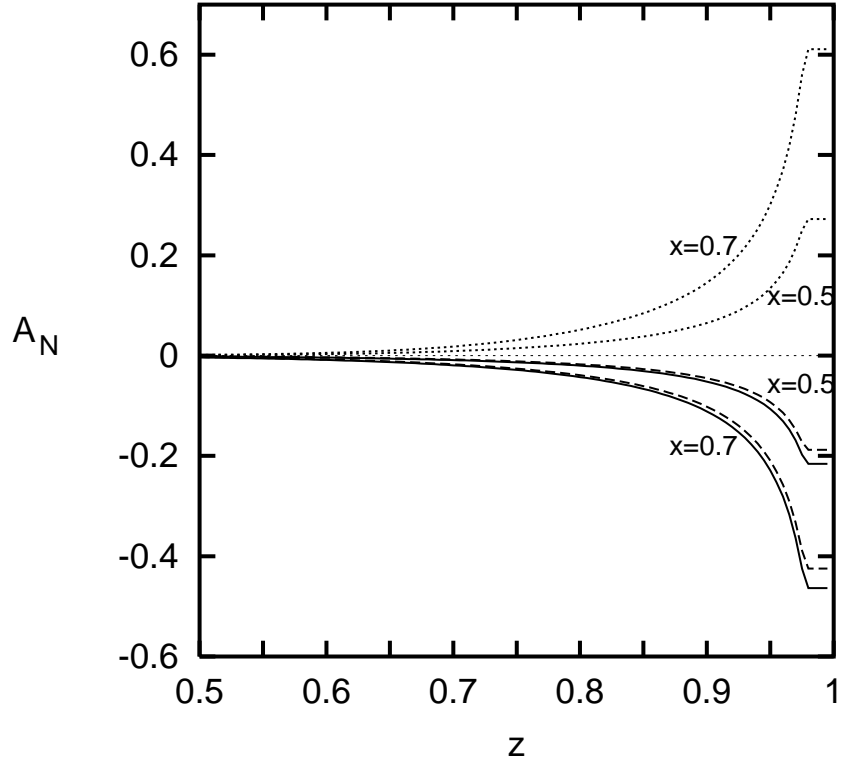,angle=-90,width=12cm}}
\vspace{1cm}
\caption{Prediction for the single spin asymmetry $A_N/|\cos2\beta|$,
Eq. (\ref{ancgr}) of text,  
for pion production in the process $\gamma^* p^{\uparrow} \to \pi X$ as a 
function of $z$. Here we have taken $Q^2=5.0$ GeV$^2$ and  
$s=12$ GeV$^2$. The two different sets of curves correspond to either 
$x=0.5$ or $x=0.7$. The solid and dashed lines in the negative plane 
are for $\pi^+$ and $\pi^0$ respectively. The dotted curves in the positive 
plane correspond to $\pi^-$. 
For $z \ge 0.97742$ the single spin asymmetry is constant because of
the constraint $\Delta ^N D = -2D$ \protect\cite{noi3}. The $z$-range 
has been restricted to $0.5-1.0$ for aesthetic reason only ($A_N$ is 
zero between $z=0.0$ and $z=0.5$). Notice how the single spin asymmetry for 
$\pi^0$ is almost the same as that for $\pi^+$.}
\end{center}
\end{figure}
%%%%%%%%%%%%%%%%%%%%%%%%%%%%%%%%%%%%%%%%%%%%%%%%%%%%%%%%%%%%%%%%%%%%%%%%%%%

\newpage

%%%%%%%%%%%%%%%%%%%%%%%%%%%%%%% Fig. 8 %%%%%%%%%%%%%%%%%%%%%%%%%%%%%%%%%%%%
\begin{figure}[t]
\begin{center}
\mbox{~\epsfig{file=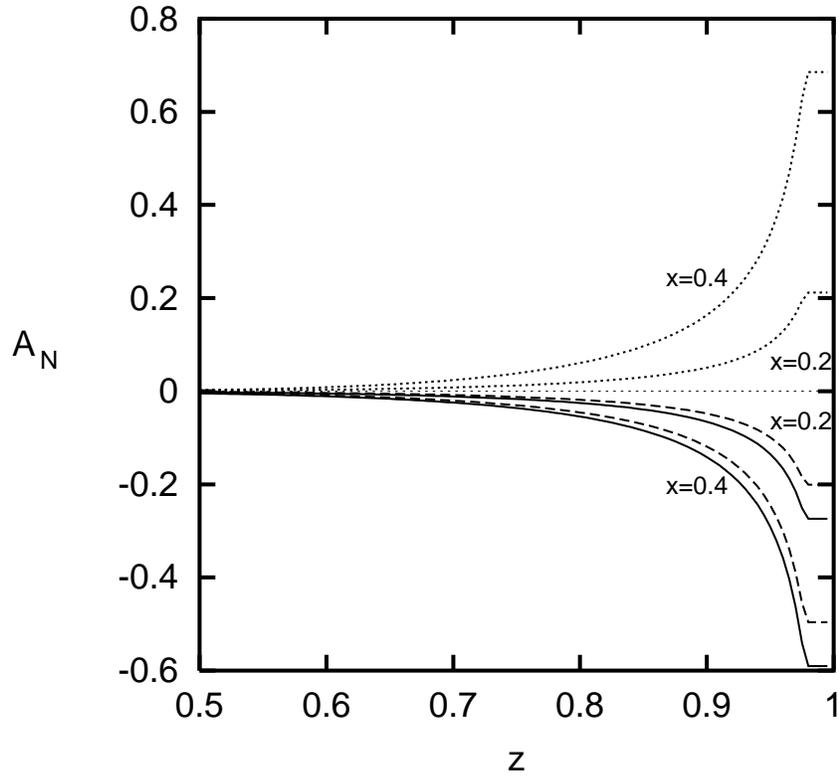,angle=-90,width=12cm}}
\vspace{1cm}
\caption{The same as in Fig.~7 with 
%Single spin asymmetry for pion production in the process 
%$\gamma ^*  p^{\uparrow} \to \pi X$ as a function of $z$. 
%Here we have taken the 
typical HERMES kinematical values: $Q^2=8.0$ GeV$^2$ and  
$s=52.4$ GeV$^2$. The two different sets of 
curves correspond to either $x=0.2$ or $x=0.4$ (notice that the cut in $x$ 
for HERMES is $0.02\le x \le 0.4$). 
%The solid and dashed lines in the negative 
%plane are for $\pi^+$ and $\pi^0$ respectively. The curves in the positive 
%plane correspond to $\pi^-$. 
%For $z \ge 0.97742$ the single spin asymmetry is constant 
%because of the constraint $\Delta ^N D = -2D$. The $z$-range has been 
%restricted to $0.5-1.0$ for aesthetic reasons only ($A_N$ is 
%zero between $z=0.0$ and $z=0.5$).
}
\end{center}
\end{figure}
%%%%%%%%%%%%%%%%%%%%%%%%%%%%%%%%%%%%%%%%%%%%%%%%%%%%%%%%%%%%%%%%%%%%%%%%%%%

\newpage 

%%%%%%%%%%%%%%%%%%%%%%%%%%%%%%% Fig. 9 %%%%%%%%%%%%%%%%%%%%%%%%%%%%%%%%%%%%
\begin{figure}[t]
\begin{center}
\mbox{~\epsfig{file=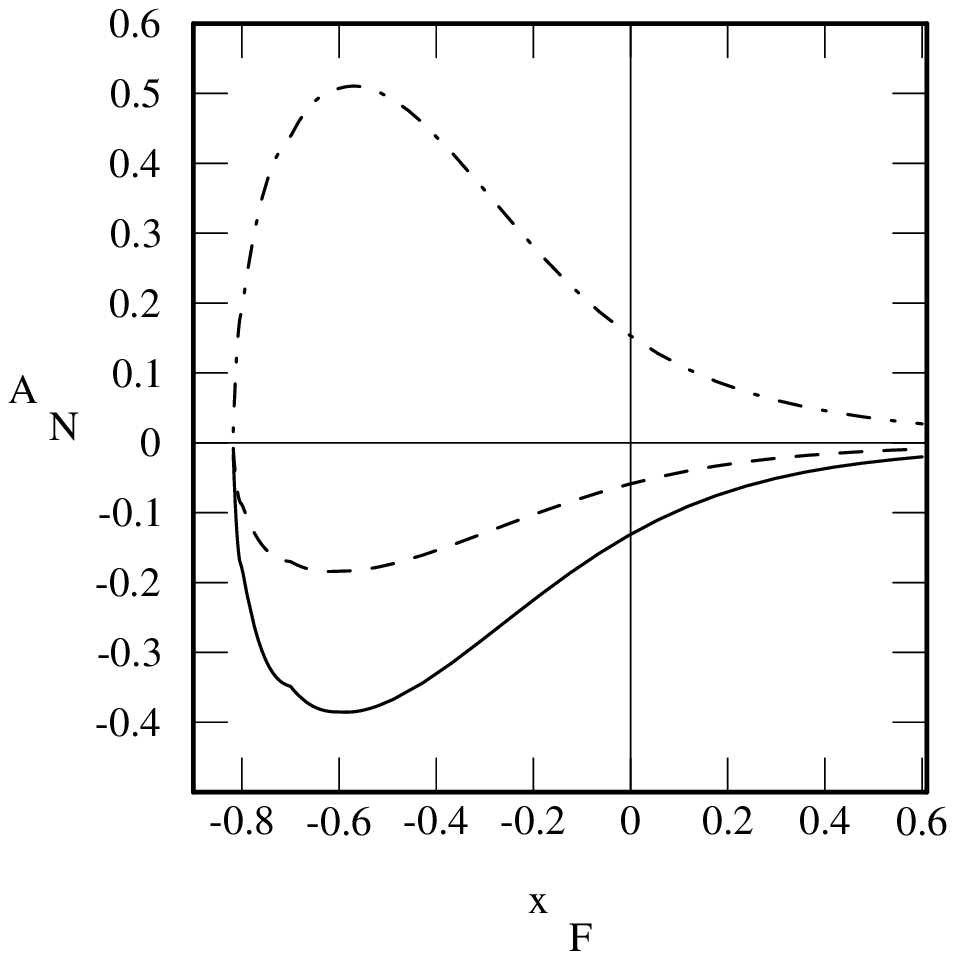,angle=0,width=12cm}}
\vspace{1cm}
\caption{Predictions for the single spin asymmetry $A_N$ in the DIS process 
$l p^{\uparrow} \to \pi X$ as a function of $x_F$: the solid line refers 
to $\pi^+$, the dashed line to $\pi^0$ and the dotted-dashed line to $\pi^-$.
We hypothetically assume that only Sivers effect is active and
the function $\Delta^N f(x,k_\perp)$ needed to calculate $A_N$ is derived from 
a fit to E704 experimental data on $p^{\uparrow} p \to \pi X$
\protect\cite{noi2}.
The value of $s$ is chosen as a planned CEBAF value, $s=12$ GeV$^2$, 
and the transverse momentum of the pion is fixed to $p_T=1.0$ GeV.}
\end{center}
\end{figure}
%%%%%%%%%%%%%%%%%%%%%%%%%%%%%%%%%%%%%%%%%%%%%%%%%%%%%%%%%%%%%%%%%%%%%%%%%%%

\newpage

%%%%%%%%%%%%%%%%%%%%%%%%%%%%%%% Fig. 10 %%%%%%%%%%%%%%%%%%%%%%%%%%%%%%%%%%%
\begin{figure}[t]
\begin{center}
\mbox{~\epsfig{file=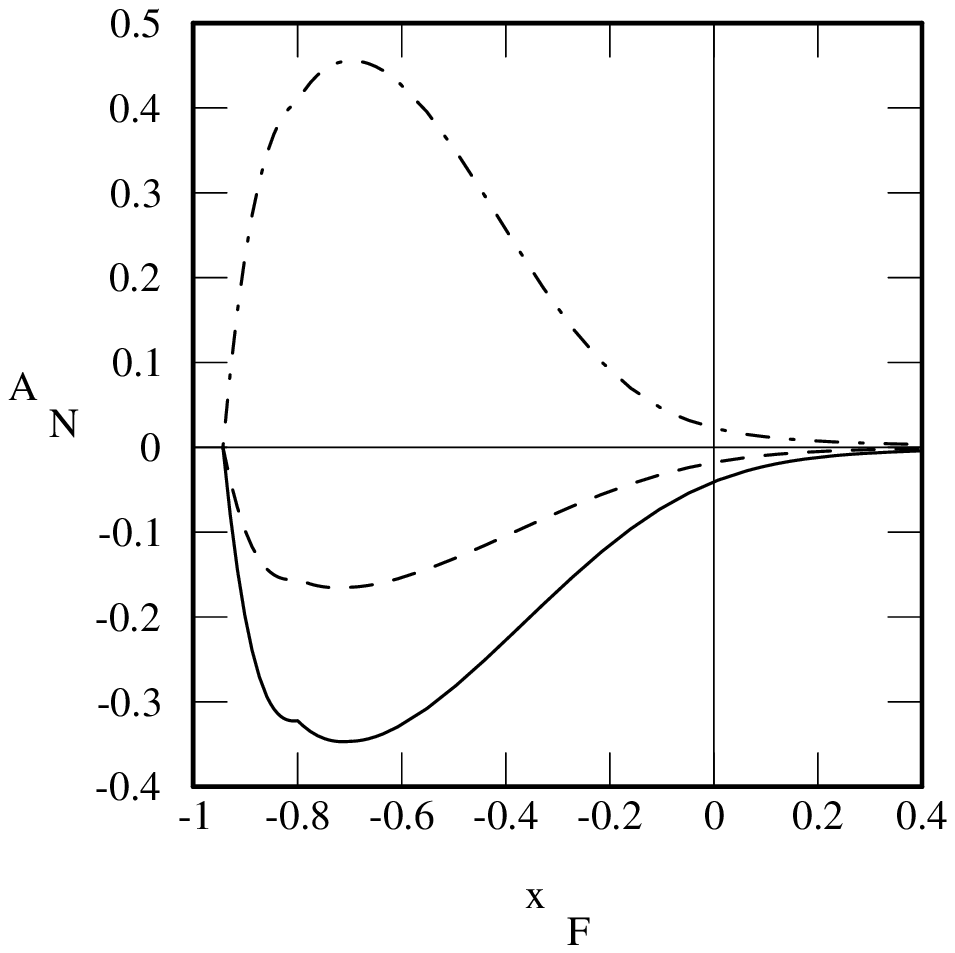,angle=0,width=12cm}}
\vspace{1cm}
\caption{The same as in Fig.~9 for typical
%Single spin asymmetry for pion production in the DIS process 
%$\l  p^{\uparrow} \to \pi X$ as a function of $x_F$.
%The solid line refers to $\pi^+$, the dashed line to 
%\pi^0$ and the dotted line to $\pi^-$.
%the function $\Delta^N D(z,k_\perp)$ needed to calculate $A_N$ is taken from 
%the fit on E704 experimental data on $p^{\uparrow}  p \to \pi X$. 
%the value of $s$ is set to be the 
kinematical values of HERMES experiment, $s=52.5$ GeV$^2$, and $p_T=1.2$ GeV.}
\end{center}
\end{figure}
%%%%%%%%%%%%%%%%%%%%%%%%%%%%%%%%%%%%%%%%%%%%%%%%%%%%%%%%%%%%%%%%%%%%%%%%%%%

\newpage

%%%%%%%%%%%%%%%%%%%%%%%%%%%% Fig. 11 %%%%%%%%%%%%%%%%%%%%%%%%%%%%%%%%%%%%%%%
\begin{figure}[t]
\begin{center}
\mbox{~\epsfig{file=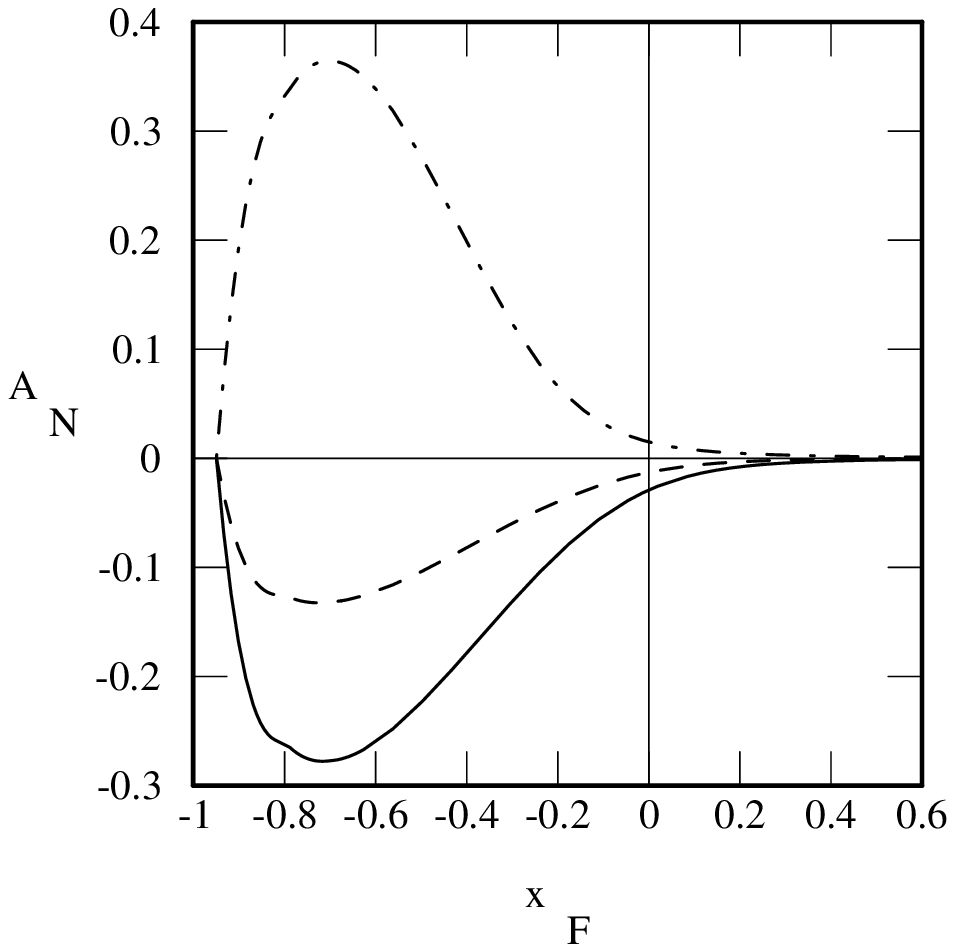,angle=0,width=12cm}}
\vspace{1cm}
\caption{The same as in Fig.~9 
%Single spin asymmetry for pion production in the DIS process 
%$l  p^{\uparrow} \to \pi X$ as a function of $x_F$.
% The solid line refers to $\pi^+$, the dashed line to 
%$\pi^0$ and the dotted line to $\pi^-$. 
%Notice that here only Collins effects are present. 
%The value of $s$ is set to be 
at SLAC energy, $s=92$ GeV$^2$, with the 
transverse momentum of the pion fixed to $p_T=1.5$ GeV.}
\end{center}
\end{figure}
%%%%%%%%%%%%%%%%%%%%%%%%%%%%%%%%%%%%%%%%%%%%%%%%%%%%%%%%%%%%%%%%%%%%%%%%%%%

\newpage

%%%%%%%%%%%%%%%%%%%%%%%%%%%%% Fig. 12 %%%%%%%%%%%%%%%%%%%%%%%%%%%%%%%%%%%%%%
\begin{figure}[t]
\begin{center}
\mbox{~\epsfig{file=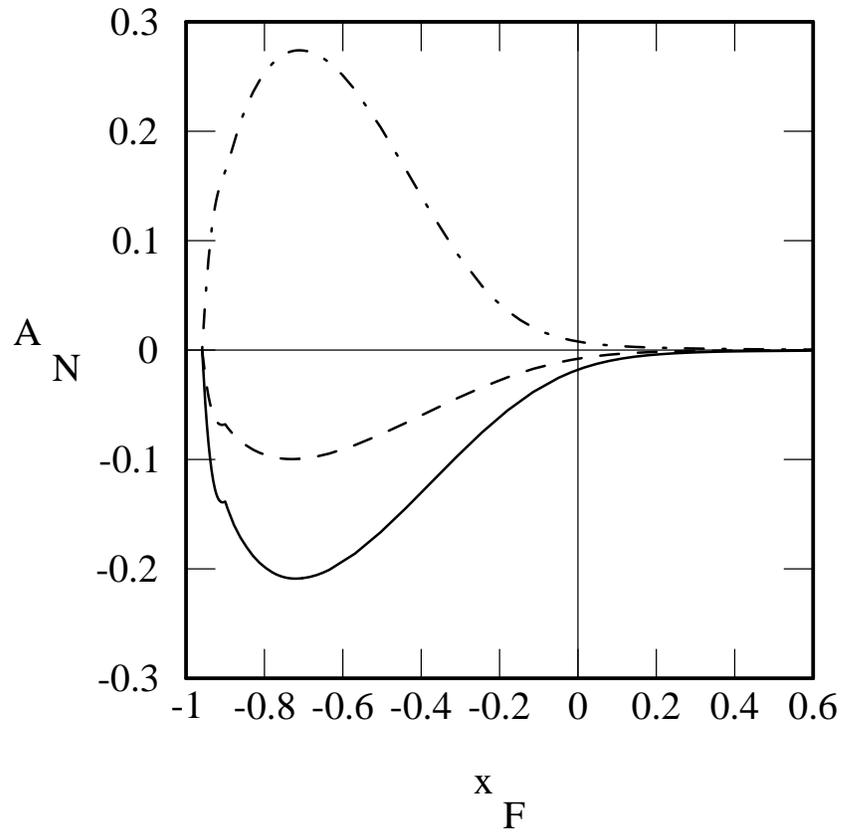,angle=0,width=12cm}}
\vspace{1cm}
\caption{The same as in Fig.~9
%Single spin asymmetry for pion production in the DIS process 
%$l  p^{\uparrow} \to \pi X$ as a function of $x_F$.
% The solid line refers to $\pi^+$, the dashed line to 
%$\pi^0$ and the dotted line to $\pi^-$. 
%Notice that here only Collins effects are present. 
%The value of $s$ is set to be the 
for the SMC energy value $s=200$ GeV$^2$, and a  
transverse momentum of the pion $p_T=2$ GeV. 
%Notice that 
%with $s=400$ GeV$^2$ and an equal pion transverse momentum $p_T$ we would 
%get a very similar result.
}
\end{center}
\end{figure}
%%%%%%%%%%%%%%%%%%%%%%%%%%%%%%%%%%%%%%%%%%%%%%%%%%%%%%%%%%%%%%%%%%%%%%%%%%%
\end{document}